\documentclass[10pt,english,twocolumn]{revtex4-2}
\usepackage[T1]{fontenc}
\usepackage[utf8]{inputenc} 
\usepackage{geometry}
\usepackage{units}
\usepackage{textcomp}
\usepackage{amsmath}
\usepackage[english]{babel}
\usepackage{graphicx}
\usepackage{xcolor}

\begin{document}


\author{E. Billaud$^1$, L. Balembois$^1$, M. Le Dantec$^1$, M. Ran\v{c}i\'{c}$^1$, E. Albertinale$^1$, S. Bertaina$^2$, T. Chaneli\`ere$^3$, P. Goldner$^4$, D. Est\`eve$^1$, D. Vion$^1$, P. Bertet$^{1*}$, E. Flurin$^1$}

\email{emmanuel.flurin@cea.fr, patrice.bertet@cea.fr}

\affiliation{$^1$Universit\'e Paris-Saclay, CEA, CNRS, SPEC, 91191 Gif-sur-Yvette Cedex, France\\
$^2$CNRS,  Aix-Marseille  Universit\'e,  IM2NP  (UMR  7334),  Institut  Mat\'eriaux Micro\'electronique  et  Nanosciences de Provence,  Marseille,  France \\
$^3$Univ. Grenoble Alpes, CNRS, Grenoble INP, Institut N\'eel, 38000 Grenoble, France \\
$^4$Chimie ParisTech, PSL University, CNRS, Institut de Recherche de Chimie Paris, 75005 Paris, France}

\title{Microwave fluorescence detection of spin echoes}

\begin{abstract}
Counting the microwave photons emitted by an ensemble of electron spins when they relax radiatively has recently been proposed as a sensitive method for electron paramagnetic resonance (EPR) spectroscopy, enabled by the development of operational Single Microwave Photon Detectors (SMPD) at millikelvin temperature. Here, we report the detection of spin echoes in the spin fluorescence signal. The echo manifests itself as a coherent modulation of the number of photons spontaneously emitted after a $\pi/2_X - \tau - \pi_Y - \tau - \pi/2_\Phi $ sequence, dependent on the relative phase $\Phi$.
We demonstrate experimentally this detection method using an ensemble of $\mathrm{Er}^{3+}$ ion spins in a scheelite crystal of $\mathrm{CaWO}_4$. We use fluorescence-detected echoes to measure the erbium spin coherence time, as well as the echo envelope modulation due to the coupling to the $^{183}\mathrm{W}$ nuclear spins surrounding each ion. We finally compare the signal-to-noise ratio of inductively-detected and fluorescence-detected echoes, and show that it is larger with the fluorescence method.
\end{abstract}

\maketitle

EPR spectroscopy is the method of choice to characterize the concentration and properties of paramagnetic impurities in a sample~\cite{schweiger_principles_2001}. For that goal, sequences of microwave pulses are applied to the spins, the most widely used being the Hahn echo ($\pi/2_X - \tau - \pi_Y - \tau - \mathrm{echo}$)~\cite{hahn_spin_1950}. Applied to an ensemble of $N$ electron spins $S=1/2$ with an inhomogeneously broadened Larmor frequency distribution, the Hahn echo sequence generates a transient build-up of macroscopic transverse magnetization $\langle S_Y \rangle$ at a time $\tau$ after the second pulse, with an amplitude that depends on multiple factors such as spin density, coherence time, nuclear spin environment, ... Therefore, spin echoes contain important information for EPR spectroscopy.

Echoes are usually detected by coupling the spins to a microwave resonator at frequency $\omega_0$; the Larmor precession of the echo transverse magnetization induces the emission into an output waveguide of a transient microwave pulse, which is then amplified and detected. Its amplitude $\langle Y_e \rangle =p N \sqrt{T_e \Gamma_r}$ depends on the average spin polarization $p$, the rate at which a spin spontaneously emits a photon into the waveguide $\Gamma_r$, and the echo duration $T_e$~\cite{bienfait_reaching_2016,albertinale_detecting_2021}. Note that because the energy stored in the spins is $p N \hbar \omega_0 / 2$, the number of echo photons $\langle Y_e \rangle^2$ cannot be larger than $p N/2$, and it is often much smaller. 
Enhancing the signal-to-noise ratio (SNR) can be achieved by using small-mode-volume and high-quality-factor resonators to increase $\Gamma_r$~\cite{narkowicz_scaling_2008,shtirberg_high-sensitivity_2011,malissa_superconducting_2013,sigillito_fast_2014,artzi_induction-detection_2015,bienfait_controlling_2016}, by cooling the sample far below $\hbar \omega_0/k_B$ to polarize the spins ($p=1$) and suppress thermal noise, and by using quantum-limited amplifiers~\cite{bienfait_reaching_2016,eichler_electron_2017,probst_inductive-detection_2017,ranjan_electron_2020}. Nevertheless, vacuum fluctuations in the echo detection mode, with standard deviation $\delta Y_e = 1/2$, ultimately impose an upper limit to the SNR achievable in this Inductive-Detection (ID) method. 

In this work, we demonstrate an alternative echo detection method, which avoids this limit and can thus reach a higher SNR than ID. We rotate the transverse magnetization generated at the echo time into a longitudinal one by adding a $\pi/2$ pulse applied with a phase $\Phi$ with respect to the first $\pi/2_X$ pulse (called the restoring pulse in the following), similar to the optical detection of spin echoes~\cite{breiland_optically_1973, oort_optically_1988}. For ideal pulses, the longitudinal polarization becomes $\langle S_Z \rangle = N/2 \cos \Phi$.
We detect this coherent modulation by counting the microwave photons emitted after the sequence when the spins relax radiatively, using a Single Microwave Photon Detector (SMPD) as demonstrated in~\cite{albertinale_detecting_2021}. Because the number of echo photons $N/2$ is much larger than $\langle Y_e \rangle ^2$, and because an ideal SMPD can detect them noiselessly, the SNR of Fluorescence-Detected (FD) echo can be much larger than in ID echo. Here, we report the observation of this FD spin-echo signal, and use it to measure the coherence time of rare-earth-ions in a crystal as well as the electron-spin-echo envelope modulation (ESEEM) due to surrounding nuclear spins. We finally compare the echo SNR reached in FD and ID in the same experimental conditions, and find it is slightly larger with FD despite the non-idealities of our experiment.

\begin{figure}
  \includegraphics[width=\columnwidth]{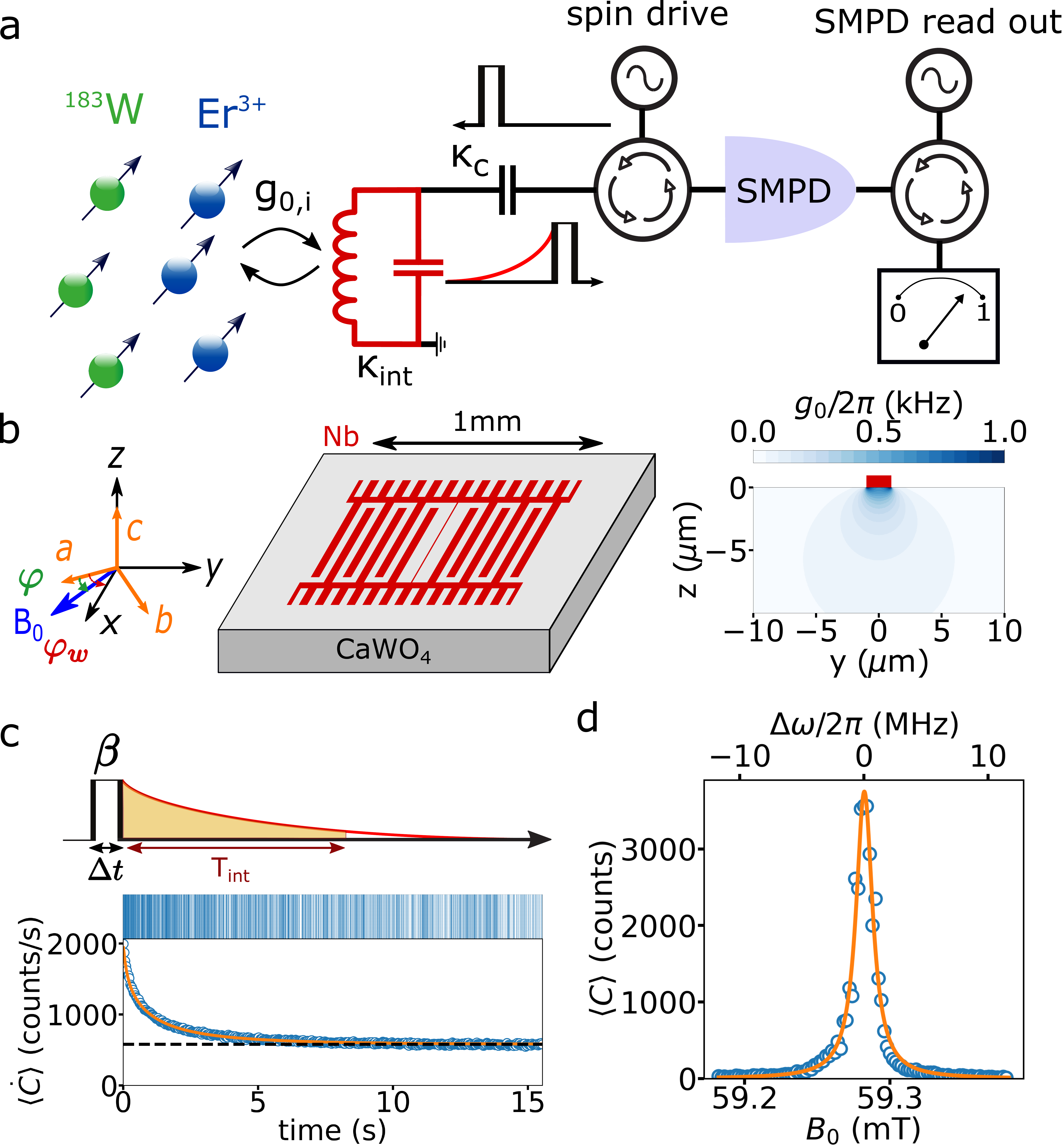}
  \caption{\label{fig1} Experiment principle and spin fluorescence signal.
  \textbf{a}, Schematic of the experiment. The resonator of frequency $\omega_0/2\pi$ is coupled to spin $i$ with a strength $g_{0,i}$. It is damped via the transmission line at a rate $\kappa_c$ and via internal losses at a rate $\kappa_{int}$.
  The transmission line is used  both for exciting the spins (black square pulses) and for routing the fluorescence signal (red decay) towards the SMPD, which is read out via another port. 
  \textbf{b}, Sketch of the niobium planar resonator fabricated on top of the $\mathrm{CaWO}_4$ sample in the $ab$ plane. The wire is along the x axis ($\varphi_w=51^{\circ}$) and $B_0$ is in the $ab$ plane ($\varphi = 31^{\circ}$).
  The cross section of the sample below the $2\ \mu m$-wide wire (shown as a red rectangle) displays the spatial distribution of the coupling $g_0$.
  \textbf{c}, Spin detection principle: a pulse of amplitude $\beta$ and duration $\Delta t$ excites the spins that then relax by emitting fluorescence photons (red curve), detected and integrated over a period $T_{int}$. Clicks (top panel-vertical bars) occur with a higher probability at short times after the excitation, hence the decay of the measured instantaneous count rate $\langle \dot{C} \rangle$ (bottom panel-circles) down to the dark count rate $\alpha = 5.8\cdot 10^{2} \ \mathrm{s}^{-1}$ (dashed line). The solid line is the simulation result. 
  \textbf{d}, EPR spectrum of $\mathrm{Er}^{3+}\mathrm{:CaWO}_4$. The average number of counts $\langle C \rangle$ (open blue circles) is shown as a function of $B_0$ (bottom axis), also converted into frequency detuning (top axis). The solid line is a fit to a Lorentzian with FWHM $1.6\ \mathrm{MHz}$. Data were obtained with SMPD1.
  }
\end{figure}

For our demonstration, we use an ensemble of erbium $\mathrm{Er}^{3+}$ ions in a scheelite crystal of $\mathrm{CaWO}_4$~\cite{mims_spectral_1961,antipin_a._paramagnetic_1968,enrique_optical_1971,bertaina_rare-earth_2007,le_dantec_twenty-threemillisecond_nodate}. The $\mathrm{Er}^{3+}$ ground state is a degenerate doublet, and behaves as an effective electron spin $S=1/ 2$ in the presence of a magnetic field $B_0$The main components of its anisotropic gyromagnetic tensor $\gamma_\parallel/2\pi = 17.45$\,GHz/T and $\gamma_\perp/2\pi = 117.3$\,GHz/T depend on whether $B_0$ is applied parallel or perpendicular to the crystal $c$-axis. 
Here we consider only the $I=0$ erbium isotopes. $\mathrm{CaWO}_4$ has a low nuclear-magnetic-moment density, since only the tungsten atoms have a stable spin-1/2 isotope ($^{183}\mathrm{W}$, present at $14\%$ natural abundance), and their gyromagnetic ratio $\gamma_n/2\pi = 1.8$\,MHz/T is low. The spin properties of paramagnetic rare-earth-ion-doped crystals were recently studied in the millikelvin regime, and shown to reach coherence times well above the millisecond~\cite{li_hyperfine_2020,le_dantec_twenty-threemillisecond_nodate,rancic_electron-spin_2022,alexander_coherent_2022}.
In this work we use the same sample as in Ref.~\cite{le_dantec_twenty-threemillisecond_nodate}, a pure scheelite crystal with residual $\mathrm{Er}^{3+}$ concentration $7\, \mu \mathrm{m}^{-3}$~\cite{erb_growth_2013}, enabling a quantitative comparison between results obtained by FD and by ID.

As shown in Fig.~\ref{fig1}a, the $\mathrm{Er}^{3+}$ ions are inductively coupled to a planar superconducting LC resonator deposited on top of the $ab$ oriented crystal surface. The resonator geometry schematically depicted in Fig.~\ref{fig1}b consists of a $2\mu\mathrm{m}$-wide wire that acts as an inductance, in parallel with an inter-digitated capacitor leading to a frequency $\omega_0 /2\pi = 6.999$\,GHz (resonator 1 in ~\cite{le_dantec_twenty-threemillisecond_nodate}). Because the magnetic field vacuum fluctuations $\mathbf{\delta B_1} (\mathbf{r})$ depend on the position $\mathbf{r}=(y,z)$ with respect to the inductance, the spin-photon coupling constant $g_0 (\mathbf{r}) = \mathbf{\delta B_1} (\mathbf{r}) \cdot \mathbf{\gamma} \cdot \langle \downarrow | \mathbf{S} | \uparrow \rangle$ is spatially inhomogeneous, reaching its largest value close to the inductance (see Fig.~\ref{fig1}b). The resonator has internal residual losses ($\kappa_{int} = 3.6 \times 10^5 \ s^{-1}$) and is capacitively coupled to the line ($\kappa_c = 2 \times 10^6 \ s^{-1}$), which yields a total energy damping rate $\kappa = \kappa_c + \kappa_{int}$.
We send through a heavily attenuated input line square-shaped excitation pulses of duration $\Delta t$, pulsation $\omega_0$ and amplitude at the resonator input $\beta$.
Each spin at location $\mathbf{r}$ are driven at the Rabi frequency $ 4 \beta \sqrt{\kappa_c} g_0(\mathbf{r})/\kappa $. The reflected pulse, together with the subsequent spin fluorescence signal, is routed via a circulator towards a SMPD based on a superconducting transmon qubit~\cite{lescanne_irreversible_2020,albertinale_detecting_2021}. The sample and SMPD are cooled at 10mK in a dilution refrigerator. The SMPD is operated cyclically, each cycle giving a click ($c=1$) or not ($c=0$). Two different SMPD devices were used in this work, with different cycle duration and dark count rates $\alpha$. For SMPD1 (described in \cite{balembois_new_nodate}), $\alpha = 5.8\cdot 10^{2} \ \mathrm{s}^{-1}$ and the cycle duration is $10.2\ \mathrm{\mu s}$. For SMPD2 (described in \cite{albertinale_detecting_2021}), $\alpha = 2\cdot 10^{3} \ \mathrm{s}^{-1}$ and the cycle duration is $8.6\ \mathrm{\mu s}$.

A typical spin fluorescence trace following an excitation pulse applied at $t=0$ is shown in Fig.~\ref{fig1}c. The output $c(t_j)$ of the SMPD cycle $j$ as a function of the cycle time $t_j$ shows an excess of photons following the pulse, which decays over a time scale of seconds. By repeating the traces $\sim 10^2$ times, we obtain the average count rate $\langle \dot{C}(t) \rangle$, shown in Fig.~\ref{fig1}c, $C(t) = \sum_{0\leq t_j \leq t} c(t_j)$ being the number of clicks between $0$ and $t$. Under the excitation pulse, each spin $i$ undergoes a Rabi rotation leading to a longitudinal magnetization $S_{Z,i}$. It then emits a photon with probability $(\Gamma_{r,i} / \Gamma_{1,i})(2S_{Z,i}+1)/2 $ at a rate $\Gamma_{1,i}$. Here, $\Gamma_{1,i} = \Gamma_{sl} + \Gamma_{r,i} $ is the total spin relaxation rate, sum of the spin-lattice rate $\Gamma_{sl}$ and of the radiative rate $\Gamma_{r,i} = \kappa g_{0,i}^2 / (\delta_i^2 + \kappa^2/4)$, $g_{0,i}$ being spin $i$ coupling constant and $\delta_i$ its detuning to the resonator~\cite{bienfait_controlling_2016}. Therefore, $\langle \dot{C}(t) \rangle = \eta \sum _i \Gamma_{r,i} \left[2\langle S_{Z,i} \rangle + 1\right]  e^{- \Gamma_{1,i} t} / 2$, where $0 \leq \eta \leq 1$ is the probability that the radiative de-excitation of an erbium ion gives rise to a SMPD click. Because of the inhomogeneity of $g_{0,i}$ (due to the spatial profile of the resonator mode) and of $\delta_i$ (due to the erbium ions inhomogeneous broadening), both the longitudinal polarization $\langle S_{Z,i} \rangle$ and the radiative relaxation rate $\Gamma_{r,i}$ are also widely distributed, explaining why the fluorescence curve in Fig.~\ref{fig1}c is non-exponential. This can be modeled using the known spin-lattice relaxation rate $\Gamma_{sl} = 0.2 ~ \mathrm{s}^{-1}$~\cite{le_dantec_twenty-threemillisecond_nodate}, the $\mathrm{Er}^{3+}$ concentration, and the $g_{0,i}$ and $\delta_i$ distributions. Simulation results ~\cite{albertinale_detecting_2021,billaud_fluorescence_nodate} quantitatively reproduce the observed fluorescence curve as seen in Fig.~\ref{fig1}c and yield an efficiency $\eta = 0.12 \pm 0.01$. This efficiency corresponds to the product of the contributions from the measured spin resonator internal losses ($\kappa_c/\kappa = 0.85$), SMPD duty cycle ($\eta_d = 0.78$), SMPD efficiency ($\eta_{SMPD}=0.45$), and of the losses between the spin device and the SMPD (inferred to be $\approx 0.5$).

\begin{figure}
  \includegraphics[width=\columnwidth]{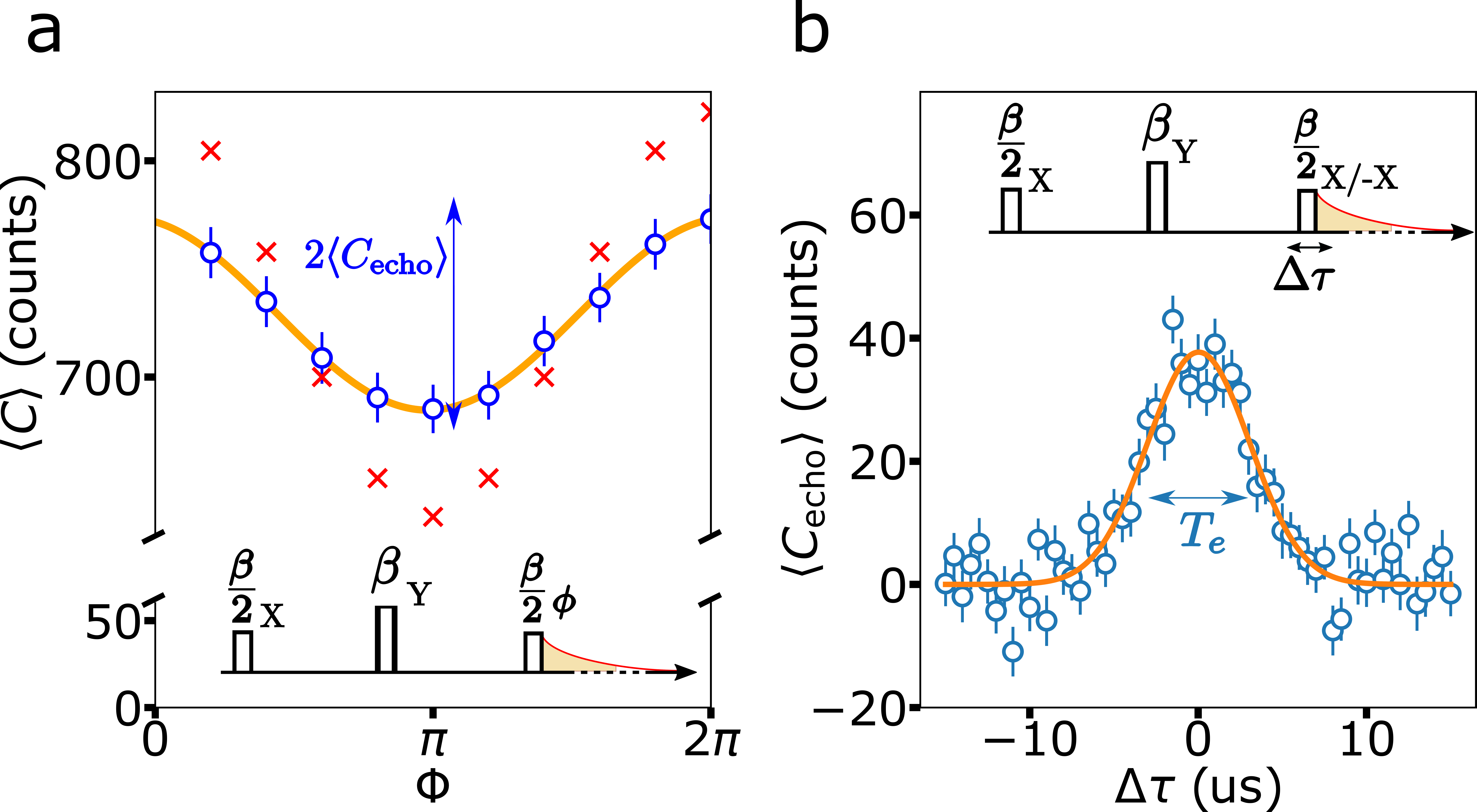}
  \caption{\label{fig2} 
  Fluorescence detection of spin echo.
  \textbf{a}, Number of counts $\langle C \rangle$ as a function of the restoring pulse phase $\Phi$. Blue dots (red crosses) are experimental (simulated) points, with $\tau = 100\ \mu \mathrm{s}$, $T_{int} = 1.52 \ \mathrm{s}$, $\beta \approx 10\ \mathrm{ns}^{-1/2}$, $\Delta t = 5 \ \mathrm{\mu s}$, and $\eta = 0.12$ (from the simulation). The solid line is a fit yielding $\langle C_\mathrm{echo} \rangle = 44$ counts and $\langle C_\mathrm{bg} \rangle = 730$ counts.
  \textbf{b}, Measured $\langle C_\mathrm{echo} \rangle$ (open blue dots) as a function of delay $\Delta \tau$ between the restoring pulse and the expected echo time. The solid line is a Gaussian fit yielding $T_e = 6.1 \mu \mathrm{s}$. Error bars are $1\sigma$ statistical. Data were obtained with SMPD1.
  }
\end{figure}

In the limit $\Gamma_{r,i} \gg \Gamma_{sl}$ and $t\gg \Gamma_{r,i}^{-1}$, the total number of counts is simply $\langle C(t) \rangle = \eta  (2\langle S_{Z} \rangle + 1)/2$, where $S_Z = \sum S_{Z,i}$ is the spin ensemble longitudinal magnetization, and is therefore the quantity of interest for spectroscopy. In practice, an integration window $T_{int}$ is defined, and the number of counts $\langle C \rangle = \langle \sum_{0\leq t_i \leq T_\mathrm{int}} c(t_i) \rangle - \alpha T_\mathrm{int}$ is computed with the dark count background subtracted. In Fig.~\ref{fig1}d, we show $\langle C \rangle(B_0)$ around $B_0 = 59$\,mT, corresponding to the erbium resonance, with the field applied at an angle $\varphi = 31^\circ$ with respect to the $a$-axis in the $ab$ plane, an angle which minimizes the erbium linewidth~\cite{mims_broadening_1966}. An approximately Lorentzian lineshape is obtained, with FWHM $\Gamma/2\pi = 1.6$\,MHz, in agreement with the results obtained by ID-EPR~\cite{le_dantec_twenty-threemillisecond_nodate}.

We now turn to the fluorescence detection of spin echoes. The sequence consists of one pulse of amplitude $\beta/2$ around the $X$ axis, followed after a delay $\tau$ by a second pulse of amplitude $\beta$ around the $Y$ axis, and after another delay $\tau$ by a third pulse of amplitude $\beta/2$ around an axis making a variable angle $\Phi$ with $X$. 
In Fig.~\ref{fig2}a, the number of counts $\langle C \rangle$ is shown as a function of $\Phi$. The data are well fitted by $\langle C(\Phi) \rangle= \langle C_\mathrm{bg} \rangle + \langle C_\mathrm{echo} \rangle \cos \Phi$, thus displaying the expected FD-echo coherent modulation.
The FD echo amplitude is measured by successively measuring the number of counts for $\Phi=0$ and $\Phi=\pi$, yielding $ C_\mathrm{echo}  \equiv [C (0) - C (\pi)]/2$. Varying the delay $\tau + \Delta \tau$ between the second and third pulses, $\langle C_\mathrm{echo} \rangle$ shows a clear echo shape with $T_e = 6.1\ \mu \mathrm{s}$ (see Fig.~\ref{fig2}b). 


Contrary to the simple model with ideal pulses, we find $\langle C_\mathrm{echo} \rangle \ll \langle C_\mathrm{bg} \rangle$, which calls for an explanation. We therefore simulate the pulse sequence, and extract $\langle C\rangle$ by performing the same analysis as with the data. The simulated $\langle C\rangle (\Phi)$, rescaled by the known efficiency $\eta$, shows an oscillation with the same phase as the data, an amplitude $\langle C_\mathrm{echo} \rangle$ approximately twice larger (see below), and a constant background value close to the measured one (see Fig.~\ref{fig2}a), indicating that the latter is due to the spread in Rabi angles among the spin ensemble, which reduces the echo amplitude as is well-known in pulsed ID-EPR. Such a spread in Rabi angle could be mitigated in future work using rapid adiabatic pulses~\cite{doll_wideband_2017, doll_adiabatic_2013,spindler_perspectives_2017,osullivan_random-access_2021}.

\begin{figure}[t!]
  \includegraphics[width=\columnwidth]{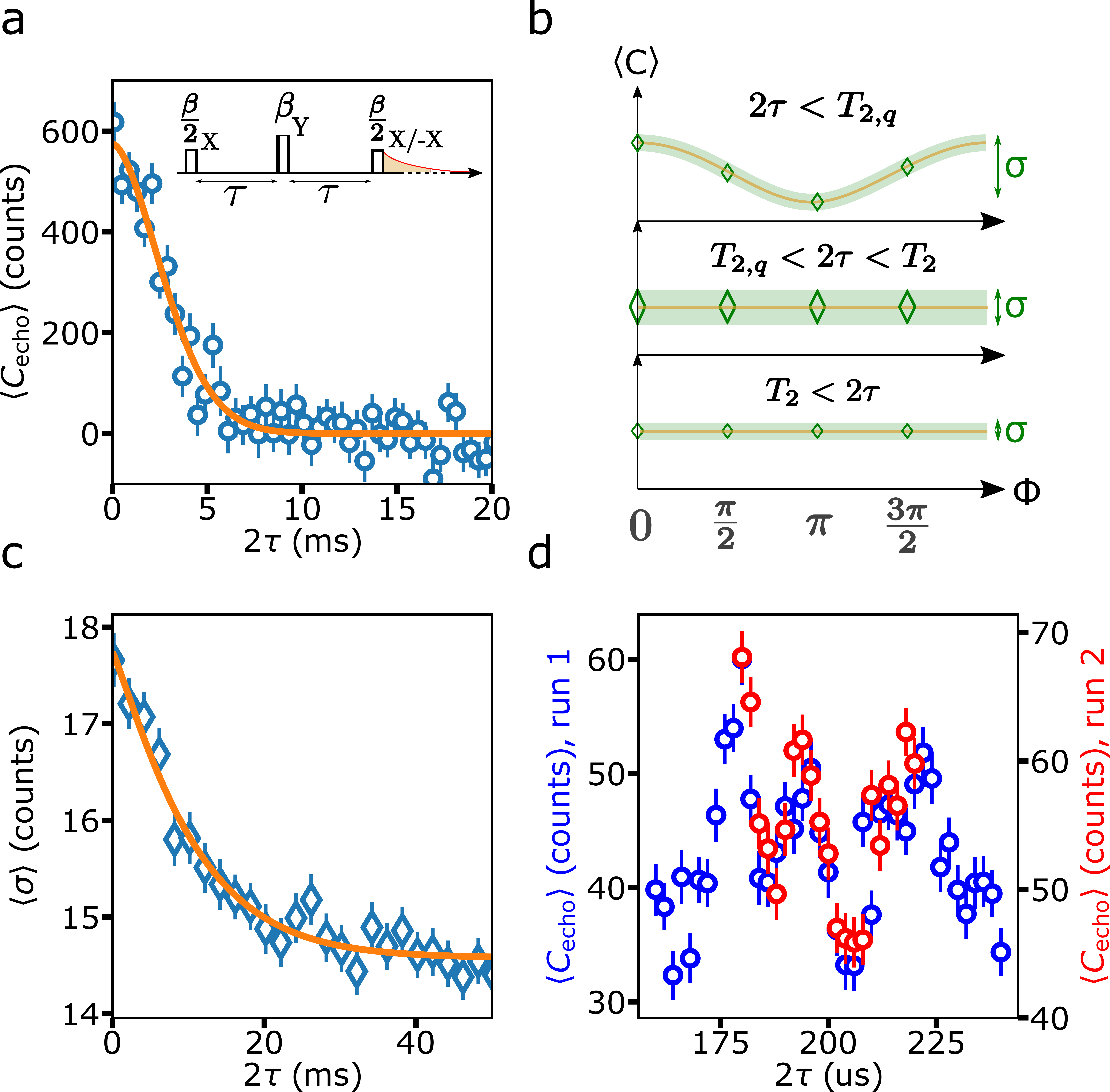}
  \caption{\label{fig3}
  Fluorescence detection of the spin coherence time
  \textbf{a}, Measured $\langle C_\mathrm{echo} \rangle$ (blue open circles) as a function of the echo delay $2\tau$, with $T_{int} = 3.7 \ \mathrm{s}$, $\beta \approx 100\ \mathrm{ns}^{-1/2}$ and $\Delta t = 5 \ \mathrm{\mu s}$. The line is a fit yielding $T_{2,q} = 3.6\,\mathrm{ms}$ and exponent $x_q=1.7$.
   \textbf{b}, Sketched dependence of the number of counts $\langle C \rangle$  as a function of the restoring pulse angle $\Phi$ at 3 different times $2\tau$.
   The orange line represents $\langle C \rangle$ and the green width the standard deviation at each angle. The diamonds show the values of $\Phi$ used in our experiment to compute the standard deviation $\sigma$, also schematically represented here. 
  \textbf{c}, Measured (open blue diamond) standard deviation $\langle \sigma \rangle$ as a function of the delay $2\tau$ (see main text). Here, $T_{int} = 0.58 \mathrm{s}$, $\beta \approx 10\ \mathrm{ns}^{-1/2}$ and $\Delta t = 5 \ \mathrm{\mu s}$. The  line is a fit yielding $T_2 = 19.8\,\mathrm{ms}$ and exponent $x=1.2$
  \textbf{d}, Measured $C_\mathrm{echo}$ as a function of delay $2\tau$, showing ESEEM. Here, $T_{int} = 0.75 \mathrm{s}$, $\beta \approx 10\ \mathrm{ns}^{-1/2}$ and $\Delta t = 5 \ \mathrm{\mu s}$. The blue and red open circles are the results from two successive runs.
  Error bars are $1\sigma$ statistical. Data were obtained with SMPD1.
  }
\end{figure}

Using the FD echo, we can measure the $\mathrm{Er}^{3+}$ spin coherence time. First, we measure $\langle C_\mathrm{echo} \rangle $ as a function of the inter-pulse delay $\tau$ (see Fig.~\ref{fig3}a). The data are fitted by a stretched exponential decay $\langle C_\mathrm{echo} \rangle = A\mathrm{e}^{-(2\tau/T_{2,q})^{x_q}}$, yielding a time constant $T_{2,q} = 3.6$\,ms and exponent $x_q=1.7$. This is consistent with the values measured in \cite{le_dantec_twenty-threemillisecond_nodate} using ID-EPR in quadrature-averaged mode.
However, in~\cite{le_dantec_twenty-threemillisecond_nodate} was found that the spin coherence time was in fact longer than $T_{2,q}$. Indeed, global magnetic field noise causes the echo transverse magnetization to acquire an extra phase $\delta \Phi$, which can be modeled as being Gaussian-distributed with a standard deviation of $ 2\tau / T_{2,q}$. Writing the number of counts obtained for a restoring pulse phase $\Phi$ as $C(\Phi,\delta\Phi,\tau) = C_\mathrm{bg} + C_\mathrm{echo} \cos (\Phi + \delta \Phi) \mathrm{e}^{-(2\tau / T_2)^x}$, $T_2$ being the spin coherence time and $x$ a streching exponent, we see indeed that $\langle C_\mathrm{echo} \rangle (\tau) = \langle C_\mathrm{echo} \rangle \mathrm{e}^{-(2\tau / T_2)^x} \mathrm{e}^{-(2\tau / T_{2,q})^2} $, thus decaying in a time $T_{2,q} \ll T_2$. 

On the other hand, it is readily shown that $\sigma \equiv \sqrt{\mathrm{Var} (C(\Phi,\delta\Phi,\tau))} \approx \sqrt{\mathrm{Var} (C_\mathrm{bg})} (1+\frac{C^2_\mathrm{echo}}{8\mathrm{Var} (C_\mathrm{bg})}\mathrm{e}^{-2(2\tau / T_2)^x})$, where the variance is taken over $\delta \Phi$ and $\Phi$ (see Fig.~\ref{fig3}b for a schematic explanation).
Experimentally, we measure the total count $C(\Phi_k,\tau)$ as a function of $\tau$ for $\Phi_k = k \pi / 2$ ($k=0,1,2,3$) and compute the standard deviation at a given $\tau$ over the 4 angles $\Phi_k$.
The data averaged over 950 repetitions (shown in Fig.~\ref{fig3}c) is fitted and yields the spin coherence time $T_2=19.8$\,ms and stretching exponent $x=1.2$. The value of $T_2$ is in good agreement with the one obtained by ID-detected echo and magnitude-averaging~\cite{tyryshkin_electron_2003,tyryshkin_electron_2012} as well as with expectations from spectral diffusion caused by the $^{183}$W nuclear spin bath dynamics~\cite{le_dantec_twenty-threemillisecond_nodate}; the fitted stretching exponent is lower, for unknown reasons.

An important application of spin echoes is to probe the local environment of paramagnetic species through pulsed hyperfine spectroscopy~\cite{schweiger_principles_2001}. We now show that this is possible also with FD, by measuring the modulation of the $\mathrm{Er}^{3+}$ echo signal caused by the proximal nuclear spin environment~\cite{mims_spectral_1961,mims_envelope_1972,mims_exchange_1990}. In Fig.~\ref{fig3}d, $\langle C_\mathrm{echo} \rangle$ is shown as a function of the inter-pulse delay $\tau$ with a $1$ $\mu\mathrm{s}$ step size. We observe a reproducible modulation, likely due to the hyperfine coupling of the erbium electron spin to the proximal $^{183}\mathrm{W}$ nuclear spins~\cite{probst_hyperfine_2020}. We note that the echo modulation also likely explains the 50\% reduction in echo amplitude of the measurements in Fig.~\ref{fig2}a with respect to the simulations (which do not take ESEEM into account), as the chosen inter-pulse delay $\tau = 100\ \mu \mathrm{s}$ in Fig.~\ref{fig2}a is close to an ESEEM minimum, as seen in Fig.~\ref{fig3}d.

\begin{figure}[t!]
  \includegraphics[width=\columnwidth]{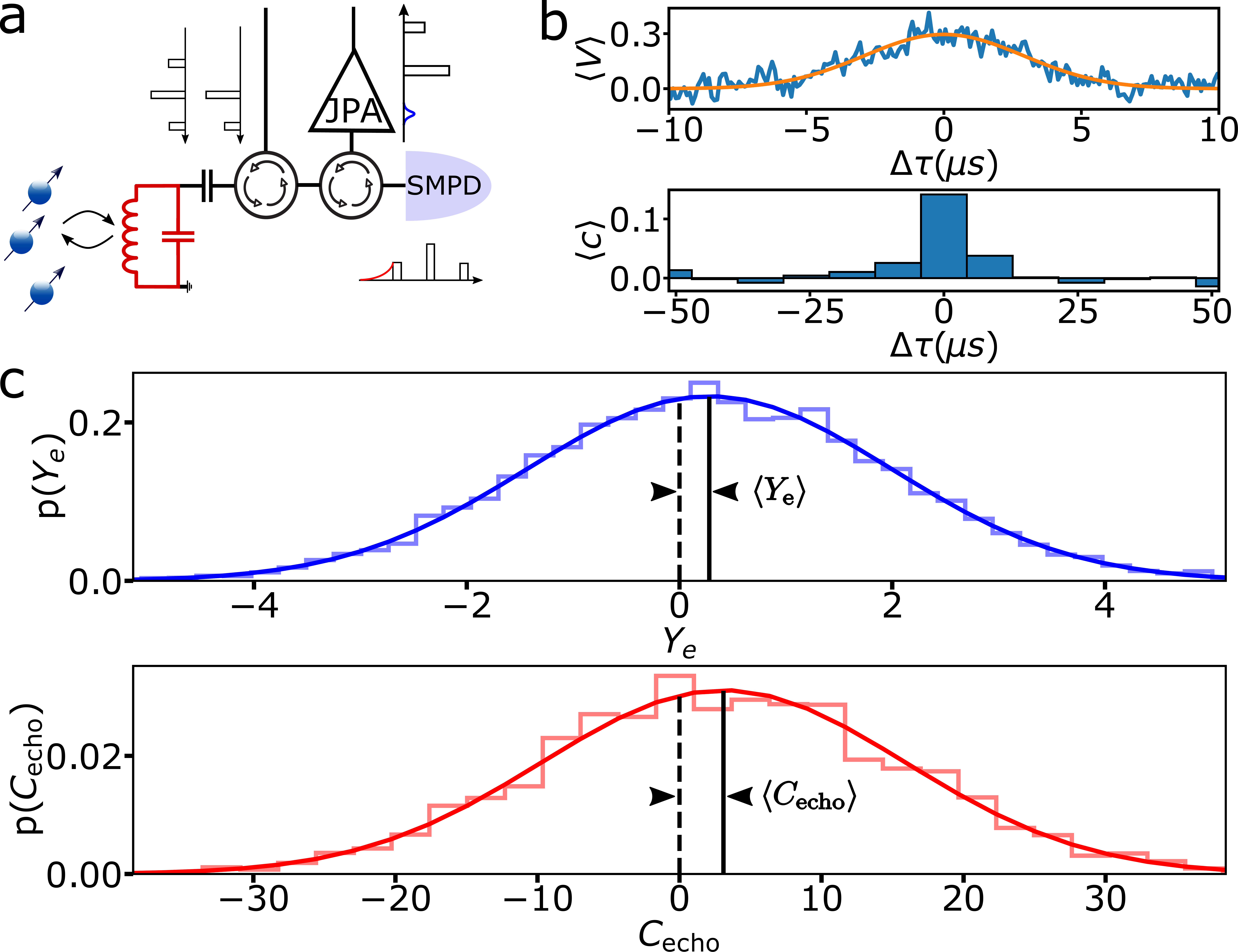}
  \caption{\label{fig4}
  SNR comparison between ID and FD detection methods
  \textbf{a}, Setup 2 used for this SNR measurement. A JPA is installed behind the SMPD2 for ID echo measurement. $B_0$ is applied with an angle $\varphi = 47^{\circ}$.
  \textbf{b}, ID and FD echo detection following the exact same two-pulse sequence with $\beta \approx 5\ \mathrm{ns}^{-1/2}$ and $\Delta t = 4 \ \mathrm{\mu s}$ to calibrate the echo photon number.
  \textbf{c}, Histograms of echo detection with FD (red, 4000 iterations) and ID (blue, 9000 iterations) taken in the same conditions, with $\beta \approx 2.5\ \mathrm{ns}^{-1/2}$ and $\Delta t = 4 \ \mathrm{\mu s}$, and their gaussian fits. Black lines show the distributions mean. Graphs are aligned on 0 in horizontal axis and are 6 standard deviation large.
  }
\end{figure}

We finally compare the SNR of FD and ID echo. For that, the setup is modified as shown in Fig.~\ref{fig4}a: a non-degenerate Josephson Parametric Amplifier (JPA) is included in the measurement chain behind the SMPD2, which makes it possible to measure ID and FD echo under the exact same conditions. We first use the SMPD to calibrate the photon number in an ID echo, and therefore the echo amplitude $Y_e$. For that, a two-pulse Hahn echo sequence is applied without a restoring pulse. A time trace of the averaged ID echo shows a Gaussian echo shape $u(t)$ (see Fig.~\ref{fig4}b). Detecting the same echo with the SMPD yields the corresponding photon number $\langle Y_e \rangle^2=0.4$, and therefore the factor needed to convert the integrated heterodyne voltage amplitude $\int V(t) u(t) dt$ into the dimensionless echo amplitude $Y_e$.

We then compare the SNR of ID two-pulse and FD restored-echo sequences, using identical first-two-pulse parameters, as well as sequence repetition time. To avoid JPA saturation, we use pulse powers much lower than in Fig.~\ref{fig1} measurements, leading to a larger relaxation rate $\Gamma_r \sim 10 \mathrm{s}^{-1}$ (data not shown). The resulting echo histograms are shown in Fig.~\ref{fig4}c. We first note that $\langle Y_e \rangle^2 = 0.08$, whereas $\langle C_{echo} \rangle = 3$, demonstrating that a larger echo signal can indeed be obtained with FD than with ID. The ratio of ID to FD echo photon numbers is $ \langle Y_e \rangle^2 / (\langle C_{echo} \rangle/\eta_d \eta_{SMPD}) = 8 \cdot 10^{-3}$, not too far off from the simple scaling $2 N \Gamma_r T_e = 2 \cdot 10^{-3}$ predicted for ideal pulses and no losses, given that $N = 2 \langle C_{echo} \rangle / (\eta_d \eta_{SMPD}) = 20$, and $T_e = 6.1 \mu \mathrm{s}$.

The standard deviation of the ID echo is found to be $\delta Y_e = 1.7$. This is close to, although larger than, the value $1/\sqrt{2}$ expected from the sole contribution of vacuum fluctuations and added noise from the quantum-limited amplifier, likely due to contributions from other amplifiers in the detection chain, as well as possible over-estimation of the conversion factor. The standard deviation of the FD echo $\delta C_{\mathrm{echo}} = 12$ counts is, on the other hand, entirely dominated by the dark count contribution. Even then, the measured FD echo signal-to-noise ratio $\mathrm{SNR}_\mathrm{FD}=\langle C_\mathrm{echo} \rangle/2\delta C_{\mathrm{echo}}=0.12$ is slightly larger than the ID echo $\mathrm{SNR}_\mathrm{ID} = \langle Y_e \rangle / 2 \delta Y_e = 0.075$. This result confirms the value of fluorescence detection in terms of sensitivity when applied to echo detection, even with the present generation of SMPD devices. Future SMPDs with lower dark count rates would yield further enhancements. For instance, SMPD1 in the same operating conditions  would yield double the signal-to-noise ratio, given the $1/\sqrt{\alpha}$ SNR scaling~\cite{albertinale_detecting_2021}. Note that this improvement was not measured with SMPD1 due to the absence of a parametric amplifier in the spin detection chain.

The detection of spin echoes using fluorescence detection completes the proof-of-principle results in~\cite{albertinale_detecting_2021} and establishes FD-EPR as an operational alternative to ID-EPR. To reach its full potential, FD-EPR requires spins in the Purcell regime, implying that they should be located within the $\sim pL$ resonator mode volume. Therefore, FD-EPR may prove useful for samples with small volumes, such as 2D materials. Finally, with larger radiative rate and lower dark count SMPD devices, FD-EPR has the potential to reach single-spin sensitivity in the near future.

\section*{Acknowledgements} 
We acknowledge technical support from P.~S\'enat, P.-F.~Orfila and S.~Delprat, and are grateful for fruitful discussions within the Quantronics group. We acknowledge IARPA and Lincoln Labs for providing a JTWPA used in the measurements. We acknowledge support from the Agence Nationale de la Recherche (ANR) through projects MIRESPIN (ANR-19-CE47-0011), DARKWADOR (ANR-19-CE47-0004), MARS (ANR-20-CE92-0041), and through the Chaire Industrielle NASNIQ (ANR-17-CHIN-0001) cofunded by Atos, and from the Région Ile-De-France through the DIM SIRTEQ (project REIMIC). This project has received funding from the European Union’s Horizon 2020 research and innovation program under Marie Sklodowska-Curie grant agreement no. 792727 (SMERC)

\bibliography{SpinEchoFluorescence}

\end{document}